\documentclass[letterpaper]{article}
\usepackage[utf8]{inputenc}

\usepackage{amsmath}
\usepackage{amsfonts}
\usepackage{amssymb}
\usepackage{amsthm}
\usepackage{authblk}
\usepackage{booktabs}
\usepackage{cite}
\usepackage{enumerate}
\usepackage{graphicx}
\usepackage[left=2cm,right=2cm,top=2cm,bottom=2cm]{geometry}
\usepackage{mathabx}
\usepackage{multirow}
\usepackage{subcaption}

\usepackage{color}

\newcommand{\diff}{\mathrm{d}}

\author{Pasquale Bosso\thanks{pasquale.bosso@uleth.ca}}
\affil{Theoretical Physics Group and Quantum Alberta, University of Lethbridge,\protect\\ 4401 University Drive, Lethbridge, Alberta, Canada, T1K 3M4\vspace{1em}}

\title{A rigorous Hamiltonian and Lagrangian analysis of classical and quantum theories with minimal
length}

\date{}

\begin{document}

\maketitle

\begin{abstract}
GUP is a phenomenological model aimed for a description of a minimal length in quantum and classical systems.
However, the analysis of problems in classical physics is usually approached preferring a different formalism than the one used for quantum systems, and \emph{vice versa}.
Potentially, the two approaches can result in inconsistencies.
Here, we eliminate such inconsistencies proposing particular meanings and relations between the variables used to describe physical systems, resulting in a precise form of the Legendre transformation.
Furthermore, we introduce two different sets of canonical variables and the relative map between them.
These two sets allow for a complete and unambiguous description of classical and quantum systems.
\end{abstract}

\section{Introduction}

A common feature of many quantum approaches to gravity is the existence of a minimal length scale \cite{Gross1988_1,Amati1989_1,Maggiore1993_1,Maggiore1993_2,Garay1995_1,Scardigli1999_1,Capozziello2000,AmelinoCamelia2002_1,Kurkov}, contradicting and requiring a modification of the Heisenberg Principle.
A phenomenological approach, known as Generalized Uncertainty Principle (GUP), has been described from different points of view.
In particular, some works focus on a quantum description of GUP considering a deformation of the canonical commutator between position and momentum operators \cite{Kempf1995_1,Ali2011_1}.
A different set of works considers a classical description of GUP through the modification of the canonical Poisson bracket between position and momentum inspired by the modified commutator in the previous point \cite{Pramanik2013_1,Pramanik2014_1}.

In previous studies, little attention has been paid to comparing Lagrangian and Hamiltonian approaches to problems with a minimal length.
An exception in the case of relativistic spin is \cite{Deriglazov}.
The two formalism are in fact usually assumed to be equivalent.
More specifically, when GUP is considered in the quantum domain, the usual tools of quantum mechanics are implemented, rooted in the Hamiltonian formalism (see, for example also \cite{Das2008_1,Bosso2017a,Gusson2018}).
On the other hand, for a description of classical systems or in quantum field theory, the machinery of the Lagrangian formalism is preferred (see, for example \cite{Kober2010,Das2016,Das2016a}).
However, the direct application of these formalisms, connected through a to-be-specified Legendre transformation, may lead to inconsistencies, as we will see below, related with the actual form of the Lagrangian and Hamiltonian functions.

To see how these inconsistencies arise, let us consider a GUP model for a three-dimensional system between the position $\vec{q}$ and the momentum $\vec{p}$ \cite{Ali2011_1}
\begin{equation}
	[q_i,p_j] = i \hbar \left\{\delta_{ij} - \gamma \delta \sqrt{p_k p_k} \left(\delta_{ij} + \frac{p_i p_j}{p_k p_k} \right) + \gamma^2 p_k p_k \left[\epsilon \delta_{ij} + (2 \epsilon + \delta^2) \frac{p_i p_j}{p_k p_k}\right]\right\}~, \label{eqn:GUP}
\end{equation}
where Einstein summation convention is used and where $\delta$, $\epsilon$, and $\gamma$ are three real parameters.
The description using three parameters, instead of two, has the practical purpose of separating the parameters defining the particular model, $\delta$ and $\epsilon$, from the parameter defining the scale at which one would expect GUP-effects to become relevant, $\gamma$.
One usually assumes $\gamma = 1/(M_\mathrm{Pl} c)$, where $M_\mathrm{Pl}$ is the Planck mass.
Furthermore, the same parametrization was proved useful in perturbation theory \cite{Bosso2017a}.
Finally, notice that \eqref{eqn:GUP} represents an effective model giving rise to a minimal length of order $\ell \sim \hbar \gamma$.
It has to be considered as an expansion up to second order in $\gamma$ of a more fundamental model derived from a full quantum theory of gravity.
Given its nature, the results of this paper, based on this effective model, have to be considered up to second order in $\gamma$.

In this framework, the ``physical'' momentum $\vec{p}$ is often expanded in terms of a ``non-physical'' one, $\vec{p}_0$, corresponding to the generator of translations
\begin{align}
	p_i = & p_{0,i} \left[1 - \gamma \delta \sqrt{p_{0,k} p_{0,k}} + \gamma^2 (\epsilon + \delta^2) p_{0,k} p_{0,k}\right]~, & 
	\mbox{with } p_{0,i} = & - i \hbar \frac{\partial}{\partial q_{0,i}} & 
	\mbox{ such that } [q_{0,i},p_{0,j}] = & i \hbar \delta_{ij}~, \label{def:p_0}
\end{align}
where $\vec{q}_0$ is the generalized coordinate operator to which $\vec{p}_0$ is the conjugate momentum.
Notice that this expansion is consistent with the modified commutator \eqref{eqn:GUP}.
The dependence on $\vec{p}_0$ only in the first relation of \eqref{def:p_0} is a direct consequence of the fact that the modified commutator depends only on the momentum.
On the other hand, considering terms depending on the position in \eqref{def:p_0} implies position-dependent terms in the modified commutator \eqref{eqn:GUP}.
These terms correspond to a minimal uncertainty in momentum, as explained in \cite{Kempf1995_1}.
We will not consider this case in this paper.
In principle, one can also assume a relation between $\vec{q}_0$ and $\vec{q} = \vec{q}(\vec{q}_0)$.
However, it is easy to prove that $\vec{q} = \vec{q}_0$ is the only relation that can fulfill \eqref{eqn:GUP} and \eqref{def:p_0}.
Being therefore equivalent, in the following we will represent the generalized coordinate by $q_i$, omitting the subscript 0.

As we stated above, some incongruences may arise when the Lagrangian and Hamiltonian approaches are compared.
Let us then consider for the moment a free system in the Hamiltonian formalism, that is as described by a Hamiltonian of the form
\begin{equation}
	H = \frac{p_i p_i}{2m}~. \label{eqn:free_hamiltonian}
\end{equation}
It is easy to show that in Heisenberg picture, one finds
\begin{subequations}
\begin{align}
	\dot{q}_i = & \frac{1}{i \hbar} [q_i,H] = \frac{p_i}{m} \left[1 - 2 \gamma \delta \sqrt{p_k p_k} + \gamma^2 (3 \epsilon + \delta^2) p_k p_k\right] = \frac{p_{0,i}}{m} \left[1 - 3 \gamma \delta \sqrt{p_{0,k} p_{0,k}} + 2 \gamma^2 (2 \epsilon + 3 \delta^2) p_{0,k} p_{0,k} \right]~, \label{eqn:q_dot_H} \\
	\dot{p}_i = & - \frac{1}{i \hbar} [p_i,H] = 0~.
\end{align}
\end{subequations}
On the other hand, when the free Lagrangian is considered
\begin{equation}
	L = \frac{1}{2} m \dot{q}_i \dot{q}_i~,
\end{equation}
one obtains for the momentum conjugate to $q_i$
\begin{equation}
	\frac{\partial L}{\partial \dot{q}_i} = m \dot{q}_i~.
\end{equation}
This last relation differs from both the expressions of $\dot{\vec{q}}$ in terms of $\vec{p}$ or $\vec{p}_0$ in \eqref{eqn:q_dot_H}, contradicting the definition of $\vec{p}_0$ as the momentum conjugate to $\vec{q}$ given in \eqref{def:p_0}.
As a consequence, the connection between the Hamiltonian $H$ and the Lagrangian $L$ is not clear.
In fact, given the problem above, it is not clear which object will take the place of the momentum in a Legendre transformation.
In what follows, we will resolve these ambiguities in the context of classical mechanics.

The paper is structured as follows:
in Section \ref{sec:Legendre}, we analyze the correct form of the Legendre transformation with a minimal length, therefore prescribing the connection between Hamilton and Lagrange formalism;
in Section \ref{sec:eom}, we study how interactions propagate from one formalism to the other, with particular attention to the minimal coupling;
in Section \ref{sec:canonical}, we introduce a canonical transformation to a new set of variables in which the physical momentum $\vec{p}$ appears as canonical momentum;
in Section \ref{sec:examples}, we illustrate two examples as applications of the results of the paper;
finally, in Section \ref{sec:conclusions}, we conclude the paper summarizing the main results and proposing the next steps in this line of research.

\section{Legendre transformation with GUP} \label{sec:Legendre}

In this section, we study the connection between Lagrangian and Hamiltonian in presence of a minimal length, sorting out the ambiguities found in the previous section.
We first introduce some definitions inspired by the usual connection between classical and quantum mechanics.
In particular, the analogy between the (quantum) Heisenberg's and the (classical) Hamilton's equations
\begin{equation}
	\dot{q}_i = \frac{1}{i \hbar} [q_i,H] \qquad \leftrightarrow \qquad \dot{q}_i = \{q_i,H\}~,
\end{equation}
is preserved with GUP only under the following definitions
\begin{align}
	\mbox{(quantum) } q_i = & i \hbar \frac{\partial}{\partial p_{0,i}} \mbox{ in $p$-representation}&
	\mbox{(classical) } \{A,B\} = & \sum_i \left(\frac{\partial A}{\partial q_i} \frac{\partial B}{\partial p_{0,i}} - \frac{\partial A}{\partial p_{0,i}} \frac{\partial B}{\partial q_i}\right)~. \label{def:Poisson}
\end{align}
In fact, these are the only definitions that imply \eqref{eqn:q_dot_H} in both quantum and classical mechanics.
The Poisson brackets $\{q_i,p_j\}$ can then be computed once the relation between $\vec{p}$ and $\vec{p}_0$ is defined.
For this purpose, we will assume the relation \eqref{def:p_0}.
From these definitions, we obtain the following fundamental Poisson brackets
\begin{subequations}
\begin{align}
	\{q_i,q_j\} = & 0~,\\
	\{p_i,p_j\} = & 0~,\\
	\{q_i,p_j\} = & \delta_{ij} - \gamma \delta \sqrt{p_k p_k} \left(\delta_{ij} + \frac{p_i p_j}{p_k p_k} \right) + \gamma^2 p_k p_k \left[\epsilon \delta_{ij} + (2 \epsilon + \delta^2) \frac{p_i p_j}{p_k p_k}\right]~.
\end{align}
\end{subequations}

To achieve a consistent description in the Lagrangian formalism, we need to assign to $\vec{p}_0$ a deeper meaning than a mathematical tool.
In fact, the requirement of $\vec{p}_0$ being the momentum conjugate to $\vec{q}$, as suggested by \eqref{def:p_0}, implies that the Hamiltonian $H$ and the Lagrangian $L$ have to be related by the following Legendre transformation \cite{Goldstein_H}
\begin{equation}
	L = \left[\dot{q}_i p_{0,i} - H\right]_{p_{0,i} = p_{0,i}(q_i,\dot{q}_i)} 
	= \frac{1}{2} m \dot{q}_i \dot{q}_i \left[1 + 2 \gamma \delta m \sqrt{\dot{q}_j \dot{q}_j} - 2 \gamma^2 (\epsilon - 3 \delta^2) m^2 \dot{q}_j \dot{q}_j \right] ~, \label{eqn:free_lagrangian}
\end{equation}
where we have used the inverse relation
\begin{equation}
	p_{0,i}(\dot{\vec{q}}) = m \dot{q}_i \left[1 + 3 \gamma \delta m \sqrt{\dot{q}_j \dot{q}_j} - 4 \gamma^2 (\epsilon - 3 \delta^2) m^2 \dot{q}_j \dot{q}_j\right]~.
\end{equation}
This implies
\begin{equation}
	\frac{\partial L}{\partial \dot{q}_i} 
	= m \dot{q}_i \left[1 + 3 \gamma \delta m \sqrt{\dot{q}_j \dot{q}_j} - 4 \gamma^2 (\epsilon - 3 \delta^2) m^2 \dot{q}_j \dot{q}_j\right] = p_{0,i}~, \label{eqn:conjugate_p_0}
\end{equation}
that is consistent with what we saw before.

We can easily find that this result can be obtained as a consequence of the form of Legendre transformation and of Hamilton's equations.
Let us then consider a generic Lagrangian $L$ and the corresponding Hamiltonian obtained through a Legendre transformation
\begin{equation}
	H = \bar{p}_i \dot{q}_i - L~, \qquad \mbox{where} ~ \bar{p}_i = \frac{\partial L}{\partial \dot{q}_i}~.
\end{equation}
In this case, $\bar{p}_i$ is just the momentum conjugate to $q_i$.
Here we do not assume any relation between $\bar{p}_i$ and $p_i$ or $p_{0,i}$.
Assuming Hamilton's equations
\begin{equation}
	\dot{q}_i = \frac{\partial H}{\partial p_{0,i}}~, \label{eqn:Hamilton_eq_q}
\end{equation}
we find
\begin{equation}
	\dot{q}_i
	= \frac{\partial^2 L}{\partial \dot{q}_j \partial p_{0,i}} \dot{q}_j + \frac{\partial L}{\partial \dot{q}_j} \frac{\partial \dot{q}_j}{\partial p_{0,i}} - \frac{\partial L}{\partial p_{0,i}}
	= \frac{\partial \bar{p}_j}{\partial p_{0,i}} \dot{q}_j~.
\end{equation}
Therefore $\frac{\partial \bar{p}_j}{\partial p_{0,i}} = \delta_{ij}$ or $\bar{p}_i \propto p_{0,i}$.
Furthermore, since in the standard theory $\bar{p}_i = p_{0,i}$, this must be the case with GUP as well.
It is furthermore worth mentioning that imposing the condition \eqref{eqn:Hamilton_eq_q} ensures that the inverse Legendre transformation has the same form as the direct one
\begin{equation}
	L = p_{0,i} \dot{q}_i - H~, \qquad \mbox{where} ~ \dot{q}_i = \frac{\partial H}{\partial p_{0,i}}~.
\end{equation}

It is interesting to note that the same Legendre transformation, when applied on the free Lagrangian
\begin{equation}
	L = \frac{1}{2} m \dot{q}_i \dot{q}_i~, \label{eqn:free_lagrangian_noGUP}
\end{equation}
results in a different Hamiltonian than \eqref{eqn:free_hamiltonian}.
In fact, in this case the Hamiltonian is
\begin{equation}
	H = \frac{p_{0,i} p_{0,i}}{2m}~. \label{eqn:free_hamiltonian_no_GUP}
\end{equation}
Hence, we are led to an ambiguity as to which approach gives the correct description of a free system:
the one described by the free Lagrangian \eqref{eqn:free_lagrangian_noGUP}, that will produce a description identical to theories without GUP;
the one described by the free Hamiltonian in \eqref{eqn:free_hamiltonian}, as it is commonly assumed when GUP is considered;
an intermediate description, different from the previous two.
At present, no direct element that could distinguish and make us prefer any of these cases is available.
Nonetheless, phenomenological arguments and sensible assumptions allow us to discard the first and the third descriptions in favour of the second.
Consider, for example, a one-dimensional quantum harmonic oscillator, described by the (quantum) Hamiltonian
\begin{equation}
	H = \frac{\bar{p}^2}{2 m} + \frac{1}{2} m \omega^2 q^2~, \label{eqn:HO_Hamiltonian_bar}
\end{equation}
where no assumption is made concerning to $\bar{p}$.
The existence of a zero point energy is usually associated with the uncertainty relation between position and momentum.
With the standard uncertainty relation, one obtains the usual zero point energy $E_0 = \hbar \omega / 2$.
On the other hand, it is easy to see that a modified uncertainty relation produces deviations from the standard zero point energy (see for example \cite{Kempf1995_1}).
Therefore, the Hamiltonian \eqref{eqn:HO_Hamiltonian_bar} with $\bar{p} = p_0$ cannot reproduce the effects of a minimal length.
However, when then relation $\bar{p} = p$ is considered, the same Hamiltonian does produce such a deviation, suggesting for a free Hamiltonian the form in \eqref{eqn:free_hamiltonian}.
Furthermore, since \eqref{eqn:free_hamiltonian} represents a free Hamiltonian, consisting of the kinetic term only, it is natural to assume that the momentum appearing in this relation is the physical momentum $p$.

Having settled the questions concerning the forms of Lagrangian and Hamiltonian for a free system, we are now going to include interactions.

\section{Equations of motion} \label{sec:eom}

Given the results of the previous section, in particular the connection between Lagrangian and Hamiltonian with a minimal length, we are going now to study the cases of interacting systems, specifically through position and velocity dependent potentials.

First, we notice that the particular relation \eqref{def:p_0}, and specifically the fact that $\vec{p}$ is proportional to $\vec{p}_0$, where the proportionality factor is a function of $|\vec{p}_0|$, implies that the physical momentum $\vec{p}$ inherits some of the properties of the generator of translations.
For example, for a free system, both $\vec{p}_0$ and $\vec{p}$ are constants of motion.
Furthermore, since we derived the Lagrangian in \eqref{eqn:free_lagrangian} starting from the free Hamiltonian in \eqref{eqn:free_hamiltonian}, we can interpret this particular Lagrangian as the kinetic energy of a system described by the generalized coordinate $\vec{q}$.
However, when a potential is considered, a straightforward application of the Legendre transformation above allows for finding the Lagrangian of the system starting from the Hamiltonian and \emph{vice versa}.

Let us consider the particular case of a conservative, position-dependent force $\vec{F}(\vec{q})$.
We find that $\dot{\vec{p}}_0$ is given by the usual Newton's second law, while in terms of the physical momentum the equations of motion are
\begin{equation}
	\dot{p}_{0,i}
	= \dot{p}_j \left[\delta_{ij} + \gamma \delta \sqrt{p_k p_k} \left(\delta_{ij} + \frac{p_i p_j}{p_k p_k}\right) - \gamma^2 (\epsilon - \delta^2) p_k p_k \left(\delta_{ij} + 2 \frac{p_i p_j}{p_k p_k} \right) \right] = F_i(\vec{q})~,
\end{equation}
where we used the relation
\begin{equation}
	p_{0,i} = p_i[1 + \gamma \delta \sqrt{p_j p_j} - \gamma^2 (\epsilon - \delta^2) p_j p_j]~.
\end{equation}
This result can be found using either Euler--Lagrange or Hamilton's equations.

As for the case of a velocity and position dependent potential, let us consider the following Lagrangian
\begin{equation}
	L = \frac{1}{2} m \dot{q}_i \dot{q}_i \left[1 + 2 \gamma \delta m \sqrt{\dot{q}_j \dot{q}_j} - 2 \gamma^2 (\epsilon - 3 \delta^2) m^2 \dot{q}_j \dot{q}_j \right] - V(\vec{q}, \dot{\vec{q}})~,
\end{equation}
consisting of the terms in \eqref{eqn:free_lagrangian}, considered as the kinetic terms, and a potential $V(\vec{q}, \dot{\vec{q}})$.
Defining $P_{0,i} = p_{0,i} + \partial V / \partial \dot{q}_i$, we then find
\begin{align}
	p_{0,i} = & m \dot{q}_i \left[1 + 3 \gamma \delta m \sqrt{\dot{q}_j \dot{q}_j} - 4 \gamma^2 (\epsilon - 3 \delta^2) m^2 \dot{q}_j \dot{q}_j \right] - \frac{\partial V}{\partial \dot{q}_i}~.\\
	\dot{q}_i = & \frac{P_{0,i}}{m} \left[1 - 3 \gamma \delta \sqrt{P_{0,j} P_{0,j}} + 2 \gamma^2 (2 \epsilon + 3 \delta^2) P_{0,j} P_{0,j} \right]~, \\
	H = P_{0,i} \dot{q}_i - \frac{\diff V}{\diff \dot{q}_i} \dot{q}_i - L = &
	\frac{1}{2} \frac{P_{0,i} P_{0,i}}{m} \left[1 - 2 \gamma \delta \sqrt{P_{0,j} P_{0,j}} + \gamma^2 (2 \epsilon + 3 \delta^2) P_{0,j} P_{0,j} \right] - \frac{\partial V}{\partial \dot{q}_i} \dot{q}_i + V(\vec{q}, \dot{\vec{q}}) \nonumber\\
	= & \frac{1}{2} \frac{P_i P_i}{m} - \frac{\partial V}{\partial \dot{q}_i} \dot{q}_i + V~,
\end{align}
where
\begin{equation}
	P_i = P_{0,i} \left[1 - \gamma \delta \sqrt{P_{0,j} P_{0,j}} + \gamma^2 (\epsilon + \delta^2) P_{0,j} P_{0,j} \right]~.
\end{equation}
Notice that this last relation is equivalent to \eqref{def:p_0}.
Furthermore, we see that
\begin{equation}
	\dot{p}_{0,i} = - \frac{\partial H}{\partial q_i}
	= - \frac{\partial V}{\partial q_i}~,
\end{equation}
that is, also in this case, the evolution of $\vec{p}_0$ is given by the same equations of standard classical mechanics.

\subsection{Electromagnetic interaction}

As an example, consider the case of a vector potential $\vec{A}$ and a scalar potential $\phi$.
In our notation we have $V(\vec{q}, \dot{\vec{q}}) = - e \left[A_i(\vec{q}) \dot{q}_i - \phi(\vec{q})\right]$.
It is then interesting to observe that with GUP the minimal coupling consists of the usual substitution $p_{0,i} \rightarrow P_{0,i} = p_{0,i} - e A_i$.
The Hamiltonian then becomes
\begin{equation}
	H = \frac{P_i P_i}{2 m} + e \phi~.
\end{equation}
This substitution preserves the equations of motion under a gauge transformation.
Furthermore, when it is considered in the quantum description, the state covariantly transforms under a gauge transformation, acquiring a phase in the exact same way it does in the standard theory.
Notice that this is possible only when the operator $\vec{p}_0$ is coupled with the vector potential $\vec{A}$ as described above.
Any other prescription fails in covariantly transforming a physical state under gauge transformation.
Similar results have also been found in \cite{Das2009_1}.

It is interesting to note that, using the Hamiltonian above, the time derivative of $\vec{p}_0$ is identical to what one would find in the standard theory
\begin{equation}
	\dot{p}_{0,i} = e \frac{\partial A_j}{\partial q_i} \dot{q}_j - e \frac{\partial \phi}{\partial q_i} = e \left[ \left(\dot{\vec{q}} \times \vec{B}\right)_i + \frac{\diff A_i}{\diff t} + E_i \right]~,
\end{equation}
in agreement with the statements above, where we introduced the magnetic and electric fields
\begin{align}
	B_i = & \epsilon_{ijk} \frac{\partial A_k}{\partial q_j}~, &  E_i = & - \frac{\partial A_i}{\partial t} - \frac{\partial \phi}{\partial q_i}~.
\end{align}
Modifications due to GUP appear when the acceleration of a particle in an electromagnetic field is computed.
In fact, we find
\begin{equation}
	m \ddot{q}_i = e \left[\left(\dot{\vec{q}} \times \vec{B}\right)_i + E_i \right] \left[ \delta_{ij} - 3 \gamma \delta m \sqrt{\dot{q}_k \dot{q}_k} \left( \delta_{ij} + \frac{\dot{q}_i \dot{q}_j}{\dot{q}_k \dot{q}_k}\right) + \gamma^2 (4 \epsilon - 3 \delta^2) m^2 \dot{q}_k \dot{q}_k \left( \delta_{ij} + \frac{\dot{q}_i \dot{q}_j}{\dot{q}_k \dot{q}_k}\right) \right]~.
\end{equation}
Notice that for the parameters $\delta, \epsilon \rightarrow 0$ we recover the usual Lorentz force.

Thus, the scheme that we developed in this paper consistently describes both Lagrangian and Hamiltonian with interactions in classical mechanics.

\section{Different choice of canonical variables} \label{sec:canonical}

These arguments show that we need to consider the set of variables formed by $\vec{q}$ and $\vec{p}_0$ as canonical variables in the Hamiltonian formalism rather than the set formed by $\vec{q}$ and $\vec{p}$.
The physical momentum $\vec{p}$ can then be defined as a function of $\vec{p}_0$.
Analogously, one can define a new quantity, $\vec{q'}$, as a function of $\vec{q}$ and $\vec{p}_0$, that will have the role of a position to which the momentum $\vec{p}$ is the conjugate momentum.
To find the actual form of $q'_i$, consider the following generating function (see Appendix \ref{apx:generating_function})
\begin{equation}
	F(\vec{p}_0, \vec{q'}) = - q'_i p_{0,i} \left[1 - \gamma \delta \sqrt{p_{0,k} p_{0,k}} + \gamma^2 (\epsilon + \delta^2) p_{0,k} p_{0,k}\right]~. \label{eqn:generating_function}
\end{equation}
We then find \cite{Goldstein_H}
\begin{subequations}\label{eqns:transformations}
\begin{align}
	p_i = - \frac{\partial F}{\partial q'_i} = & p_{0,i} \left[1 - \gamma \delta \sqrt{p_{0,k} p_{0,k}} + \gamma^2 (\epsilon + \delta^2) p_{0,k} p_{0,k}\right]~, \\
	q_i	= - \frac{\partial F}{\partial p_{0,i}} = & q'_j \left\{\delta_{ij} - \gamma \delta \sqrt{p_k p_k} \left( \delta_{ij} + \frac{p_i p_j}{p_k p_k} \right) + \gamma^2 p_k p_k \left[ \epsilon \delta_{ij} + (2 \epsilon + \delta^2) \frac{p_i p_j}{ p_k p_k} \right] \right\}~, \label{eqn:q_of_q'}\\
	q'_i = & q_j \left\{ \delta_{ij} + \gamma \delta \sqrt{p_{0,k} p_{0,k}} \left( \delta_{ij} + \frac{p_{0,i} p_{0,j}}{p_{0,k} p_{0,k}} \right) - \gamma^2 p_{0,k} p_{0,k} \left[\epsilon \delta_{ij} + (2 \epsilon - \delta^2) \frac{p_{0,j} p_{0,i}}{ p_{0,k} p_{0,k}} \right] \right\}~.
\end{align}
\end{subequations}
This transformation is a canonical transformation, and since $F$ does not depend on time, the Hamiltonian in the new set of variables is the same as the Hamiltonian in the old set.
Furthermore, notice that
\begin{equation}
	\{q'_i,p_j\}_{q,p_0} = \delta_{ij}~, \label{eqn:Poisson_new}
\end{equation}
where we indicated by a subscript the variables with respect to which the Poisson brackets are defined.
On the other hand, since the above transformation is canonical, the Poisson brackets are invariant
\begin{equation}
	\{A,B\}_{q',p} = \frac{\partial A}{\partial q'_i} \frac{\partial B}{\partial p_i} - \frac{\partial B}{\partial q'_i} \frac{\partial A}{\partial p_i}
	= \{A,B\}_{q,p_0}~,
\end{equation}
as it can also be directly shown.
Therefore, we also have
\begin{equation}
	\{q'_i,p_j\}_{q',p} = \delta_{ij}~.
\end{equation}
It is then obvious that the equation of motion can be easily written choosing the appropriate set of coordinates.
In particular, we see that
\begin{align}
	\dot{q}_i = \frac{\partial H}{\partial p_{0,i}} = & \left\{q_i,H\right\}_{q,p_0}~, &
	\dot{p}_i = - \frac{\partial H}{\partial q'_i} = & \left\{p_i,H\right\}_{q',p}~. \label{eqn:eom}
\end{align}
In other words, what we called physical position and momentum, $\vec{q}$ and $\vec{p}$, respectively, belong to two distinct canonical sets of phase space variables, connected through a canonical transformation generated by the function $F$ in \eqref{eqn:generating_function}.
This feature is common in realizations of GUP, allowing also a direct connection to Doubly Special Relativity theories \cite{Galan2007}.

As a consequence, this new set of variables are relevant since now the free Hamiltonian \eqref{eqn:free_hamiltonian} is written directly in terms of canonical variables.
On the other hand, when we apply a Legendre transformation, we easily obtain a Lagrangian in terms of $\vec{q'}$ in the usual form for a free system
\begin{equation}
	L = \frac{1}{2} m \dot{q'}_i \dot{q'}_i~.
\end{equation}
Since $\vec{p}$ is the momentum conjugate to $\vec{q'}$, for a free system we find
\begin{equation}
	p_i = m \dot{q'}_i~,
\end{equation}
that, using the relations in \eqref{eqns:transformations}, reduce to \eqref{eqn:q_dot_H}.

Finally, it is worth noticing that, in quantum mechanics, the relation \eqref{eqn:q_of_q'} gives the correct form of the position operator in momentum space described in \cite{Kempf1995_1}.
More generically, promoting $\vec{q'}$ and $\vec{p}$ to operators, with $[q'_i,p_j] = i\hbar$, as suggested by \eqref{eqn:Poisson_new}, the relation \eqref{eqn:q_of_q'} furnishes the appropriate representation of the position operator in momentum space when the model in \eqref{eqn:GUP} is considered.
Furthermore, it is then possible to construct the formal eigenfunctions for the position operator.
For example, in one dimension we can easily find
\begin{equation}
	\psi_\lambda (p) = \sqrt{\frac{\gamma \sqrt{3 \epsilon}}{\pi}} \exp \left\{-i \frac{\lambda}{\hbar \gamma \sqrt{3 \epsilon}} \arctan \left[\frac{\gamma (3 \epsilon + \delta^2) p - \delta}{\sqrt{3 \epsilon}}\right] \right\}~.
\end{equation}
Note that this wavefunction resembles the one in \cite{Kempf1995_1} for $\delta=0$ and $\epsilon = 1/3$.

\section{Examples} \label{sec:examples}

In this section, we will illustrate two examples, showing how the results of this paper, and in particular the definition of the new set of canonical variables, allows for solving problems in classical mechanics with a minimal length.

\subsection{the Harmonic Oscillator}

A typical example that is usually worked out as an application of similar analysis is the one concerning a one-dimensional harmonic oscillator.
Following this tradition, consider the customary Hamiltonian
\begin{equation}
	H = \frac{p^2}{2m} + \frac{1}{2} m \omega^2 q^2~.
\end{equation}
In terms of the pair of variables $q$ and $p_0$, we have
\begin{align}
	H = & \frac{p_0^2}{2m} \left[1 - 2 \gamma \delta p_0 + \gamma^2 (2 \epsilon + 3 \delta^2) p_0^2\right] + \frac{1}{2} m \omega^2 q^2~. \label{eqn:hamiltonian_HO}
\end{align}
Using relations \eqref{eqn:eom}, we easily obtain
\begin{align}
	\dot{q} = & \frac{p}{m} \left[1 - 2 \gamma \delta p + \gamma^2 (3 \epsilon + \delta^2) p^2 \right]~, &
	\dot{p} = & - m \omega^2 q \left[ 1 - 2 \gamma \delta p + \gamma^2 (3 \epsilon + \delta^2) p^2 \right]~. \label{eqn:HO_eom}
\end{align}
Combining these two equations together, we find
\begin{equation}
	\ddot{q} = - \omega^2 q \left[1 - 6 \gamma \delta p + 12 \gamma^2 (3 \epsilon + \delta^2) p^2 \right]~.
\end{equation}
As for the other two canonical variables, we find
\begin{align}
	\dot{p_0} = & - m \omega^2 q~, &
	\dot{q'} = & \frac{p}{m} - 2 m \omega^2 q'{}^2 \left[\gamma \delta - 3 \gamma^2 (\epsilon + \delta^2) p\right]~.
\end{align}

We now want to solve this problem using Hamilton-Jacobi method, that is, introducing a generating function for a canonical transformation $S(p_0,\alpha,t)$ such that the new Hamiltonian vanishes identically \cite{Goldstein_H}.
Here, $\alpha$ is, by definition, a constant of motion.
The position $q$ is related to $S$ via
\begin{equation}
	q = - \frac{\partial S}{\partial p_0}~.
\end{equation}
Substituting it into \eqref{eqn:hamiltonian_HO}, we have the following partial differential equation
\begin{equation}
	\frac{p^2}{2m} + \frac{1}{2} m \omega^2 \left(\frac{\partial S}{\partial p_0}\right)^2 = - \frac{\partial S}{\partial t}~,
\end{equation}
where we used $p$ instead than $p_0$ to have a more compact notation.
However note that, since we are considering a description in terms of $q$ and $p_0$, the quantity $p$ has to be considered as a function of $p_0$.
We now consider the following \emph{ansatz} for Hamilton's principal functions
\begin{equation}
	S(p_0,\alpha,t) = W(p_0,\alpha) - \alpha t~, 
\end{equation}
motivated by the fact that only the RHS of the previous differential equation acts on the explicit time dependence of $S$.
We then find as a solution up to second order in $\gamma$
\begin{multline}
	S(p_0,\alpha,t) = \mp \frac{1}{m \omega} \left\{ \sqrt{2 \alpha m - p^2} \left[ \frac{p}{2} - \frac{2}{3} \gamma \delta (2 \alpha m - p^2) + \frac{3}{4} \gamma^2 (\epsilon - \delta^2) p (\alpha m - p^2) \right] \right.\\
	\left. + \alpha m \arcsin \left(\frac{p}{\sqrt{2 \alpha m}}\right) \left[ 1 - \frac{3}{2} (\epsilon - \delta^2) \gamma^2 \alpha m \right] \right\} - \alpha t~.
\end{multline}
Imposing the following boundary conditions
\begin{align}
	q (t=0) = & 0~, & p_0(t=0) = & p_{0,0} ~ \Leftrightarrow ~ p(t=0) = p_{,0}~,
\end{align}
one obtains
\begin{equation}
	\alpha = \frac{p_{,0}^2}{2m}~,
\end{equation}
that is, $\alpha$ corresponds to the initial energy.
We then find
\begin{equation}
	q = \pm \frac{\sqrt{p_{,0}^2 - p^2}}{m \omega} = \pm \frac{p_{,0}}{m\omega} \sqrt{1 - x^2}~, \qquad \mbox{where}~ x = \frac{p}{p_{,0}}~.
\end{equation}

As for the variable conjugate to $\alpha$, we have
\begin{equation}
	\beta = t + \frac{\arcsin(x)}{\omega} - \frac{2}{\omega} \gamma \delta p_{,0} \sqrt{1-x^2} \gamma - \frac{3}{2 \omega} (\epsilon - \delta^2) p_{,0}^2 \left(\arcsin x - x \sqrt{1-x^2}\right) \gamma^2~. \label{eqn:beta}
\end{equation}
This is a constant of motion by definition as well, and can be used to find the time dependence of $p$.
From the boundary conditions, we find
\begin{equation}
	\beta = \frac{\pi}{4 \omega} \left[2 - 3 (\epsilon - \delta^2) p_{,0}^2 \gamma^2 \right]~.
\end{equation}
Using this expression and inverting relation \eqref{eqn:beta}, we obtain to second order in $\gamma$
\begin{equation}
	p = p_{,0} \left\{\cos (\omega t) + 2 \gamma \delta p_{,0} \sin^2 (\omega t) - \frac{3}{2} \gamma^2 \left\{2 \delta^2 \sin (2 \omega t) + \left(\epsilon - \delta^2\right) \left[\sin (2 \omega t) + \omega t \right] \right\} p_{,0}^2 \sin (\omega t)\right\}~. \label{eqn:HO_momentum}
\end{equation}
It is interesting to observe that this last equation has a term proportional to $t$, therefore $x$ results to be unbounded, while from \eqref{eqn:beta} $x$ is constrained between -1 and 1.
This is due to the fact that the last equation is a product of an expansion in $\gamma$, and therefore cannot be considered reliable for large momenta compared to the Planck momentum and/or for large time intervals compared to the oscillation period.
For small momenta and/or for small time intervals, in fact, the contribution of the term proportional to $t$ is negligible.
It is also worth observing that in the linear and quadratic model, such that $\delta = \epsilon = 1$, the same term does not appear.
Finally, for the position we find
\begin{equation}
	q = \frac{p_{,0}}{m \omega} \left\{\sin (\omega t)
 	- \gamma \delta p_{,0} \sin (2 \omega t)
 	+ \gamma^2 \left[ 3 (\epsilon - \delta^2) \omega t \cos (\omega t) + 3 (\epsilon + \delta^2) \cos (\omega t) \sin (2 \omega t) - 4 \delta^2 \sin (\omega t) \right] p_{,0}^2 \right\}~. \label{eqn:HO_position}
\end{equation}

\begin{figure}
	\begin{subfigure}[t]{0.98\textwidth}
	\center
	\includegraphics[width=0.98\textwidth]{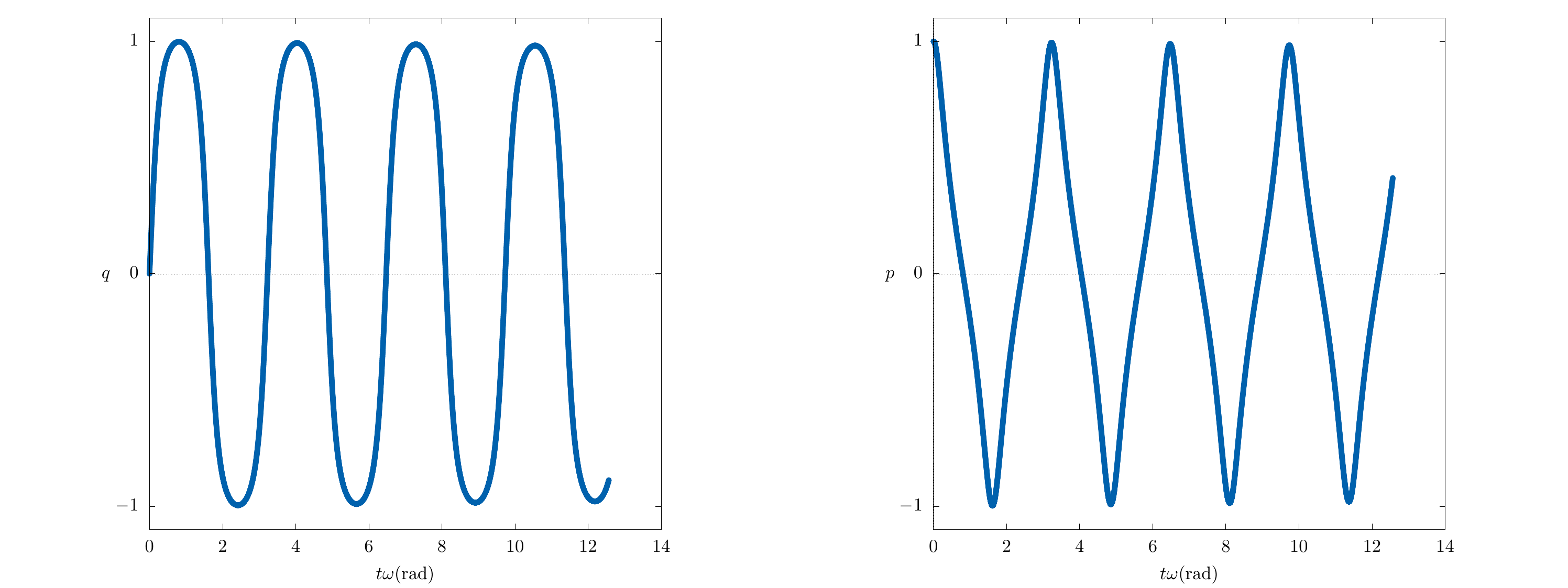}
	\caption{Time-dependence of position (left) and momentum (right) for a quadratic GUP model ($\delta = 0$, $\epsilon = 1/3$) \cite{Kempf1995_1} as obtained from \eqref{eqn:HO_eom}.
	They describe the motion of a particle of mass $m=1$Kg, oscillating with an angular frequency of $\omega=1$Hz and an initial momentum of $p_{,0}=10$Ns.
	The vertical axes represent the position and momentum in units of $p_{,0}/m \omega$ and $p_{,0}$, respectively.
	Two aspects are noteworthy: the two functions are no longer sinusoidal;
	The period of the oscillation reduces, becoming less than $T = 2 \pi / \omega$.}
	\end{subfigure}
	
	\begin{subfigure}[t]{0.98\textwidth}
	\center
	\includegraphics[width=0.98\textwidth]{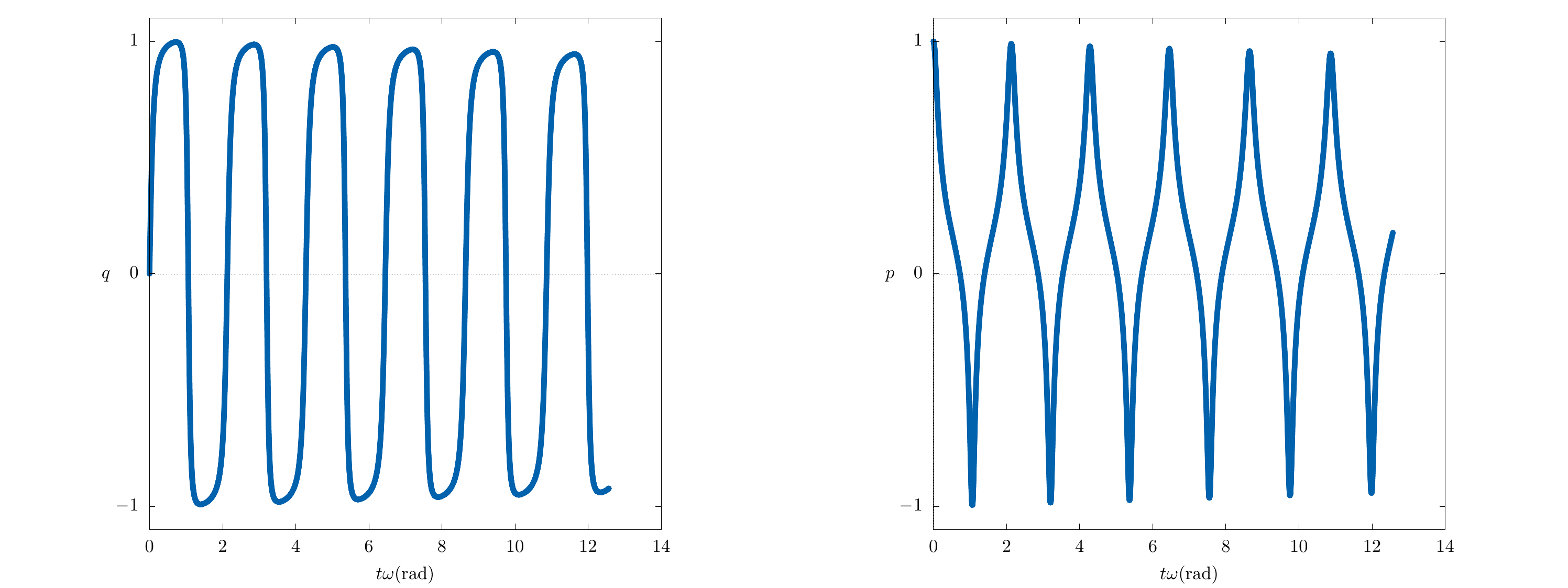}
	\caption{Time-dependence of position (left) and momentum (right) for a linear and quadratic GUP model ($\delta = 1$, $\epsilon = 1$) \cite{Ali2011_1} as obtained from \eqref{eqn:HO_eom}.
	The same parameters and axes of the previous pictures are used.
	Also in this case, the shape of the functions is changed and the period is reduced.}
	\end{subfigure}
	
	\begin{subfigure}[t]{0.98\textwidth}
	\center
	\includegraphics[width=0.5\textwidth]{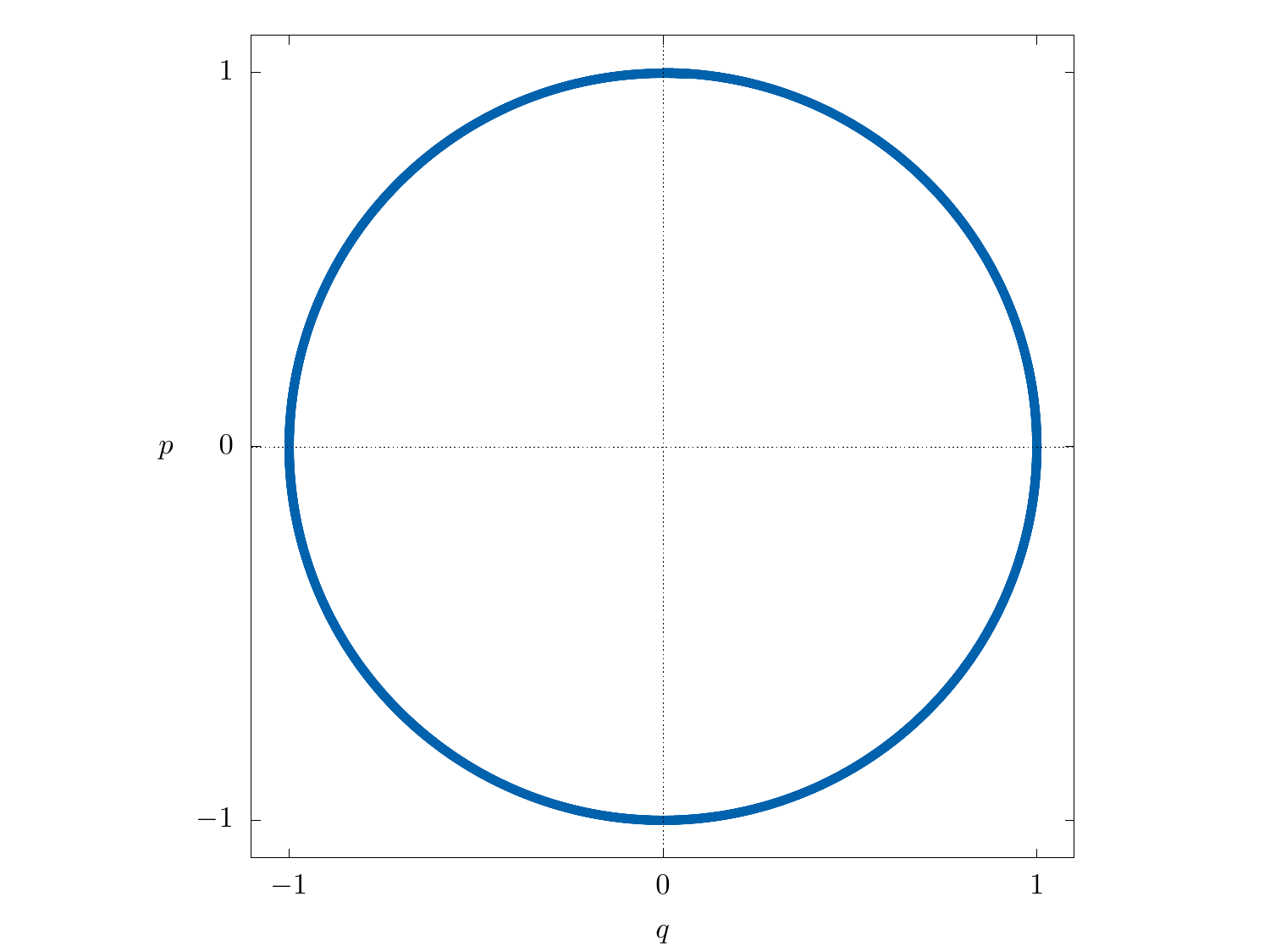}
	\caption{Phase-space orbit of a harmonic oscillator with GUP.
	The $x$- and $y$- axes represent the position and momentum in units of $p_{,0}/m \omega$ and $p_{,0}$, respectively.
	Both quadratic and linear$+$quadratic models give the same figure, independently of the property of the particle.} \label{fig:phase-space_HO}
	\end{subfigure}
	\caption{}
\end{figure}
It is interesting to notice that, both numerically integrating \eqref{eqn:HO_eom} or considering equations \eqref{eqn:HO_momentum} and \eqref{eqn:HO_position} for small values of $p_{,0}$, one obtains the usual phase-space orbit for a harmonic oscillator, as shown in Fig. \ref{fig:phase-space_HO}.
In this picture, a particle of mass $m=1$Kg, oscillating with an angular frequency of $\omega=1$Hz and an initial momentum of $p_{,0}=10$Ns is considered.
Although for these values we have $p_{,0} \gamma = 1.67$, that is a momentum for the point mass comparable with the Planck momentum, no evidence of Planck effects appear in the phase-space diagram.

\subsection{Kepler's Problem}

Another interesting problem that can be described using the formalism in this paper is Kepler's problem.
Consider a two-body system, described by a total mass $M$, a reduced mass $m$, a relative position $\vec{q}$, and a relative momentum $\vec{p}$.
Since the potential does not depend on the velocities of the particles, to derive the Hamiltonian from the Lagrangian, we simply need the functional relation between velocities and momenta.
To find this relation, we can focus on the kinetic part of the Lagrangian, given in \eqref{eqn:free_lagrangian}.
In spherical coordinates, we then find
\begin{align}
	p_{0,r} = & m \dot{r} A~, & p_{0,\theta} = & m r^2 \dot{\theta} A~, & p_{0,\phi} = & m r^2 \sin^2 \theta \dot{\phi} A~,
\end{align}
where $r$, $\theta$, $\phi$ are the radius, the colatitude and longitude angles, respectively, while
\begin{equation}
	A = \frac{\partial L}{\partial (\dot{q}_i \dot{q}_i)} = 1 + 3 \gamma \delta m \sqrt{\dot{q}_i \dot{q}_i} - 4 \gamma^2 (\epsilon - 3 \delta^2) m^2 \dot{q}_i \dot{q}_i~,
\end{equation}
where the Lagrangian has the following form
\begin{equation}
	L = \frac{1}{2} m \dot{q}_i \dot{q}_i \left[1 + 2 \gamma \delta m \sqrt{\dot{q}_i \dot{q}_i} - 2 \gamma^2 (\epsilon - 3 \delta^2) m^2 \dot{q}_i \dot{q}_i \right] + \frac{G M m}{r}~,
\end{equation}
The Lagrangian does not depend on $\phi$, therefore the conjugate momentum $p_{0,\phi}$ must be a constant of motion.
We also see that
\begin{equation}
	\dot{q}_i \dot{q}_i = \frac{1}{m^2 A^2} \left[p_{0,r}^2 + \frac{p_{0,\theta}^2}{r^2} + \frac{p_{0,\phi}^2}{r^2 \sin^2 \theta}\right]~.
\end{equation}
From this relation, we can find $\dot{q}_i \dot{q}_i$ as a function of the momenta only.
In fact, up to second order in $\gamma$, we have
\begin{equation}
	\dot{q}_i \dot{q}_i = \frac{1}{m^2} \Pi^2 \left[ 1 - 6 \gamma \delta \Pi + \gamma^2 \left(21 \delta^2 + 8 \epsilon\right) \Pi^2 \right]~,
\end{equation}
where
\begin{equation}
	\Pi^2 = p_{0,r}^2 + \frac{\Pi_\Omega {}^2}{r^2}~, \qquad \mbox{with} \qquad \Pi_\Omega {}^2 = p_{0,\theta}^2 + \frac{p_{0,\phi}^2}{\sin^2 \theta}~.
\end{equation}
As a consequence, $A$ can be written as
\begin{equation}
	A = 1 + 3 \gamma \delta \Pi - \gamma^2 (4 \epsilon - 3 \delta^2) \Pi^2~.
\end{equation}
The Hamiltonian of this system is
\begin{equation}
	H = \frac{\Pi^2}{2 m} \left[1 - 2 \gamma \delta \Pi + \gamma^2 (2 \epsilon + 3 \delta^2) \Pi^2\right] - \frac{G M m}{r}~.
\end{equation}
From Hamilton's equations, we find
\begin{equation}
	\dot{p}_{0,\theta} = \frac{p_{0,\phi}^2}{m \sin^3 \theta} \left[ 1 - 3 \gamma \delta \Pi + 2 \gamma^2 (3 \delta^2 + 2 \epsilon) \Pi^2 \right] \cos \theta~.
\end{equation}
The momentum $p_{0,\theta}$ can then be made a constant of motion considering the boundary conditions $\theta (t=0) = \pi/2$ and $p_{0,\theta} (t=0) = 0$.
As for the radial motion, we have
\begin{align}
	\dot{r} = & \frac{p_{0,r}}{m} \left[ 1 - 3 \gamma \delta \Pi + 2 \gamma^2 (3 \delta^2 + 2 \epsilon) \Pi^2 \right]~, &
	\dot{p}_{0,r} = & \frac{p_{0,\phi}^2}{m r^3} \left[ 1 - 3 \gamma \delta \Pi + 2 \gamma^2 (3 \delta^2 + 2 \epsilon) \Pi^2 \right] - \frac{G M m}{r^2}~. \label{eqn:hamilton_r}
\end{align}
It is interesting to observe that, potentially, $\dot{r}$ can be reduced to zero without vanishing $p_{0,r}$.
On the other hand, one should notice that such a case corresponds to $\Pi \sim 1/\gamma$, that is a momentum of the order of the Planck momentum.
This scale is outside the range of validity of the current phenomenological model, for in this range higher order contribution have to be taken into account.

\begin{figure}
	\center
	\includegraphics[width=0.8\textwidth]{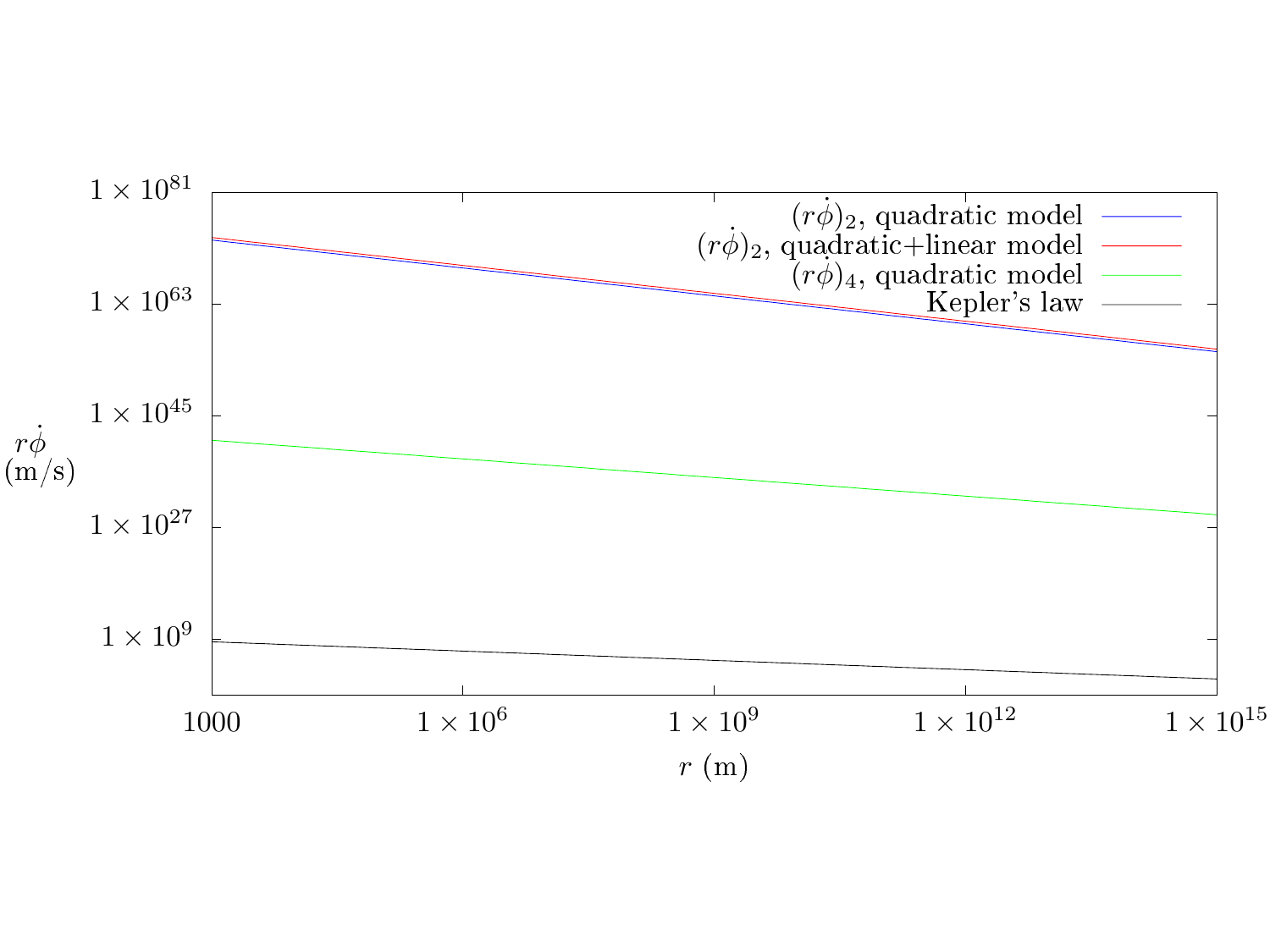}
	\vspace{-5em}
	\caption{Tangential velocity for circular orbits according to \eqref{eqns:tangential_velocity} and to Kepler's law.
	For this plot, we considered a body of 1 Earth mass moving about another body of 1 solar mass.} \label{fig:tangential_velocity}
\end{figure}
For a circular orbit, we have $\dot{r} = 0$ and $p_{0,r} = 0$.
Writing the second equation in \eqref{eqn:hamilton_r} with these conditions, we find
\begin{equation}
	r^2 \dot{\phi}^2 \left[1 + 3 \gamma \delta m r \dot{\phi} - 4 \gamma^2 (\epsilon - 3 \delta^2) m^2 r^2 \dot{\phi}^2\right] = \frac{G M}{r}~.
\end{equation}
In general, this equation will have two real roots and two conjugate complex roots for the tangential velocity $r \dot{\phi}$, that are
\begin{subequations} \label{eqns:tangential_velocity}
\begin{align}
	(r \dot{\phi})_1 = & - \frac{\sqrt{G M}}{\sqrt{r}} \left( 1 + \frac{3 \sqrt{G M} m \delta \gamma}{2\sqrt{r}} - \frac{G M m^2 (3 \delta^2 - 16 \epsilon) \gamma^2}{8 r} \right)~,\\
	(r \dot{\phi})_2 = & \frac{\sqrt{G M}}{\sqrt{r}} \left( 1 - \frac{3 \sqrt{G M} m \delta \gamma}{2\sqrt{r}} - \frac{G M m^2 (3 \delta^2 - 16 \epsilon) \gamma^2}{8 r} \right)~,\\
	(r \dot{\phi})_3 = & \frac{3 \delta + \sqrt{16 \epsilon - 39 \delta^2}}{8 m ( \epsilon - 3 \delta^2) \gamma} 
	+ \gamma \frac{G M m}{r} \frac{ 3 (16 \epsilon - 39 \delta^2) \delta - \sqrt{16 \epsilon - 39 \delta^2} (8 \epsilon - 15 \delta^2)}{2 (16 \epsilon - 39 \delta^2)}~, \\
	(r \dot{\phi})_4 = & \frac{3 \delta - \sqrt{16 \epsilon - 39 \delta^2}}{8 m ( \epsilon - 3 \delta^2) \gamma} 
	+ \gamma \frac{G M m}{r} \frac{ 3 (16 \epsilon - 39 \delta^2) \delta + \sqrt{16 \epsilon - 39 \delta^2} (8 \epsilon - 15 \delta^2)}{2 (16 \epsilon - 39 \delta^2)}~.
\end{align}
\end{subequations}
The dependence on $r$ of the relations in \eqref{eqns:tangential_velocity} has been shown in Fig. \ref{fig:tangential_velocity} for a body of 1 Earth mass.
We see that all the solutions differ from observed velocities on the Solar System scale by many orders of magnitude.
Notice, though, that we assumed $\gamma = 1/ (M_\mathrm{Pl} c) \simeq 0.15 (\mathrm{Ns})^{-1}$.
On the other hand, regarding $\gamma$ as a free parameter, comparing our results with observed velocities, and considering only real and positive solutions, a parameter roughly 30 orders of magnitude smaller than the one assumed here should be considered, corresponding therefore to a minimal length 30 orders of magnitude smaller than the Planck length.
Furthermore, the last two solutions are real only for models such that $|\delta| < 4 \sqrt{\frac{\epsilon}{39}}$, excluding the case $\epsilon = 3 \delta^2$.
These solutions correspond to finite velocities even for $r\rightarrow \infty$, while for finite radii they decay faster than the Keplerian velocity for increasing distance from the source of the field.
Notice that in the standard limit, \emph{i.e.} $\delta,\epsilon \rightarrow 0$, these solutions diverge.
In other words, in this limit these two solutions are associated with diverging velocities.
This is obviously not acceptable, especially considering the role of the speed of light as a limit velocity.

We then see that these results pose serious questions concerning models with a minimal length in classical systems.
A possible solution is in the actual value of the parameter $\gamma$, as we saw above.
On the other hand, when this phenomenological problem is considered from the point of view of fundamental constituents, the Planck scale effects enormously reduces with no change in the value of $\gamma$, because of the so-called soccer ball problem \cite{AmelinoCamelia2013_1}.
Furthermore, a better analysis has to be looked for in the relativistic description of the Kepler's problem.
As for the particular problem of the two last solutions in \eqref{eqns:tangential_velocity}, it is worth to recall that they are the result of approximated relations expanded in series of the parameter $\gamma$.
Therefore, more accurate theories may justify or eliminate the corresponding problems.
We postpone further analysis on them to future studies.

To derive the equation of motions for the physical momenta, it is convenient to change variables and to describe the Hamiltonian in terms of $\vec{q'}$ and $\vec{p}$.
In this case, the Hamiltonian reads
\begin{equation}
	H = \frac{p_i p_i}{2m} - \frac{G M m}{r(\vec{q'}, \vec{p})}~, \label{eqn:Hamiltonian_prime}
\end{equation}
where we explicitly represented $r$ as a function of $\vec{q'}$ and $\vec{p}$.
In particular, we find
\begin{equation}
	r^2 = q'_j q'_k \left\{ \delta_{jk} - 2 \gamma \delta \sqrt{p_l p_l} \left(\delta_{jk} + \frac{p_j p_k}{p_l p_l}\right) + \gamma^2 p_l p_l \left[ (2\epsilon + \delta^2) \delta_{jk} + (4\epsilon + 5\delta^2) \frac{p_j p_k}{p_l p_l}\right] \right\}~. \label{eqn:radius}
\end{equation}
Since we have
\begin{align}
	q_j' q_k' \delta_{jk} = & r' {}^2~, &
	q_j' p_j = & r' p_r~,
\end{align}
we can rewrite \eqref{eqn:radius} in spherical coordinates as
\begin{equation}
	r = r' \left\{1 - \frac{\gamma \delta}{\Pi'} \left(2 p_r^2 + \frac{\Pi'_\Omega {}^2}{r' {}^2} \right) + \gamma^2 \left[ \epsilon \left(3 p_r^2 + \frac{\Pi'_\Omega {}^2}{r' {}^2} \right) + \delta^2 \left(- p_r^2 + \frac{5}{2} \frac{\Pi'_\Omega {}^2}{r' {}^2} - \frac{\Pi'_\Omega {}^4}{r'^4 \Pi' {}^2} \right) \right]\right\}~,
\end{equation}
where
\begin{equation}
	\Pi' {}^2 = p_i p_i = p_r^2 + \frac{\Pi'_\Omega {}^2}{r'^2}~, \qquad \mbox{with} \qquad \Pi'_\Omega {}^2 = p_\theta^2 + \frac{p_\phi^2}{\sin^2 \theta'}~.
\end{equation}

We therefore see that the potential term in \eqref{eqn:Hamiltonian_prime} depends on $r'$ and $p_r$.
On the other hand, the Hamiltonian does not depend on $\phi'$.
Thus, the conjugate momentum, $p_\phi$, that is the physical angular momentum for rotations about the $z$-axis, is a constant of motion.
We also get
\begin{equation}
	\dot{\phi'} = \frac{p_\phi}{m r' {}^2 \sin^2 \theta'}~.
\end{equation}
As for the colatitude angle, we have
\begin{align}
	\dot{\theta '} = & \frac{p_\theta}{m r' {}^2}~, &
	\dot{p}_\theta = & \frac{p_\phi^2}{m r' {}^2} \frac{\cot \theta'}{\sin^2 \theta'}~.
\end{align}
We therefore see that, if we impose the initial conditions $\theta'(t=0) = \pi/2$ and $p_\theta (t=0) = 0$, then $p_\theta$ is a constant of motion.
It is worth noting that this condition is compatible with the previous one about $\theta$ and $p_{0,\theta}$.
In other words, if $p_{0,\theta}$ is a constant of motion, $p_\theta$ is a constant of motion as well.
Finally, for the radial equations of motion, we find
\begin{subequations}
\begin{align}
	\dot{r}' = & \frac{p_r}{m} - \gamma \frac{G M m}{r^2(\vec{q'}, \vec{p})} p_r r' \left\{ \frac{\delta}{\Pi'} \left(2 + \frac{\Pi'_\Omega {}^2}{\Pi' {}^2 r' {}^2} \right) - 2 \gamma \left[ 3 \epsilon - \delta^2 \frac{p_r^2}{\Pi' {}^2} \left( 1 + \frac{\Pi'_\Omega {}^2}{\Pi' {}^2 r^2} \right) \right] \right\}~,\\
	\dot{p}_r = & \frac{\Pi'_\Omega {}^2}{m r' {}^3} - \frac{G Mm}{r^2(\vec{q'}, \vec{p})} \left\{ 1 - \gamma \delta \frac{p_r^2}{\Pi' {}^3} \left(2 p_r^2 + 3 \Pi'_\Omega {}^2 \right) + \gamma^2 \left[ \epsilon \left( 3 p_r^2 - \Pi'_\Omega {}^2 \right) - \frac{\delta^2}{2} \left( \frac{2 p_r^6}{\Pi' {}^4} + 3 \Pi'_\Omega {}^2 \right) \right] \right\}~.
\end{align}
\end{subequations}
Hence, we see that both the centripetal and the gravitational forces are changed by GUP, since now they depend on the momentum $\vec{p}$, and in particular on its angular part $\Pi'_\Omega$.
Furthermore, for the specific case of a quadratic model, the radial momentum increases the gravitational force, while the angular part reduces it.
On the other hand, for a linear and quadratic model, for momenta much larger than the Planck momentum, we have a similar behavior as in the previous case, while for much lower momenta both the radial and the angular momenta reduce the gravitational force.

For a radial fall, we need to impose $\Pi'_\Omega = 0$.
This results in
\begin{align}
	\dot{r}' = & \frac{p_r}{m} - 2 \gamma \frac{G M m}{r^2} r' \left[ \delta - \gamma \left( 3 \epsilon - \delta^2 \right) p_r \right]~, &
	\dot{p}_r = & - \frac{G Mm}{r^2} \left[ 1 - 2 \gamma \delta p_r + \gamma^2 \left( 3 \epsilon - \delta^2 \right) p_r^2 \right]~.
\end{align}
This case isolates the gravitational force from the centripetal one.
Also in this case, the role of the radial momentum in the rate of change of itself is evident.

\section{Conclusions} \label{sec:conclusions}

GUP is a phenomenological model whose purpose is to account for the existence of a minimal length.
Past works have studied the influence of this feature, derived from quantum gravity theories, on quantum and classical systems.
While studies on quantum and classical effects of a minimal length have been independently pursued, finding interesting and relevant results in both frameworks, little attention has been paid on the consistency of the respective methods, rooted in Hamiltonian and Lagrangian mechanics, respectively.
In this paper we fill this gap.
In particular, we derive the necessary form of the Legendre transformation to connect Hamiltonian and Lagrangian functions with a minimal length.
Furthermore, we analyze how potentials are transformed from one formalism to the other, with particular attention to the minimal coupling.
We then define a canonical transformation to a new set of variables, in which the physical momentum $\vec{p}$ appears as canonical momentum.
Finally, we consider the results of the paper in two illustrative examples, namely the harmonic oscillator and the Kepler's problem.

Notice that, differently from \cite{Pramanik2013_1}, we do not assume any particular form of the Poisson brackets of position and momentum.
Rather, we start from the relation \eqref{def:p_0} between the physical and non-physical momenta, $\vec{p}$ and $\vec{p}_0$ respectively.
Subsequently, we show that the Poisson brackets $\{q_i,p_j\}$ resemble the quantum GUP commutator \eqref{eqn:GUP}, as it is commonly assumed.

This analysis ultimately resolves the problems concerning the choice of the formalism, making the Hamiltonian and Lagrangian approaches equivalent, and setting the stage for the study of the structure of phase space with a minimal length, necessary for quantum field theory.
Nonetheless, further study is required, in particular to extend the current results to the relativistic case.
We hope to address these aspects in future publications.

\vspace{1em}
\noindent
{\bf Acknowledgment}

\vspace{0.5em}
\noindent
I wish to thank S. Das, E. Vagenas, and V. Todorinov for interesting discussions.

\section*{Appendix}

\begin{appendix}
\section{Generating function $F(\vec{p}_0,\vec{q'})$} \label{apx:generating_function}

Consider an arbitrary function $F(\vec{p}_0, \vec{q'})$ that generates the transformation between the sets $\{\vec{q}, \vec{p}_0\}$ and $\{\vec{q'}, \vec{p}\}$ \cite{Goldstein_H}, that is
\begin{align}
	p_i = & - \frac{\partial F}{\partial q'_i}~, & q_i = & - \frac{\partial F}{\partial p_{0,i}}~.
\end{align}
The first of these equations is given by \eqref{def:p_0}.
From that expression, it is obvious that $\vec{p}_i$ cannot depend on $\vec{q'}$.
Therefore, the function $F$ is given in general by
\begin{equation}
	F(\vec{p}_0, \vec{q'}) = - q'_i p_{0,i} \left[1 - \gamma \delta \sqrt{p_{0,k} p_{0,k}} + \gamma^2 (\epsilon + \delta^2) p_{0,k} p_{0,k}\right] + K(\vec{p}_0, t)~,
\end{equation}
where $K$ is an arbitrary function of $\vec{p}_0$ and the time $t$.
We then find
\begin{subequations}
\begin{align}
	p_i = & p_{0,i} \left[1 - \gamma \delta \sqrt{p_{0,k} p_{0,k}} + \gamma^2 (\epsilon + \delta^2) p_{0,k} p_{0,k}\right]~, \\
	q_i 
	= & q'_j \left\{\delta_{ij} - \gamma \delta \sqrt{p_k p_k} \left( \delta_{ij} + \frac{p_i p_j}{p_k p_k} \right) + \gamma^2 p_k p_k \left[ \epsilon \delta_{ij} + (2 \epsilon + \delta^2) \frac{p_i p_j}{ p_k p_k} \right] \right\} - \frac{\partial K}{\partial p_{0,i}}~.
\end{align}
\end{subequations}
Inverting this last equation, we find
\begin{equation}
	q'_i
	= \left(q_j + \frac{\partial K}{\partial p_{0,j}}\right) \left\{ \delta_{ij} + \gamma \delta \sqrt{p_{0,k} p_{0,k}} \left( \delta_{ij} + \frac{p_{0,i} p_{0,j}}{p_{0,k} p_{0,k}} \right) - \gamma^2 p_{0,k} p_{0,k} \left[\epsilon \delta_{ij} + (2 \epsilon - \delta^2) \frac{p_{0,j} p_{0,i}}{ p_{0,k} p_{0,k}} \right] \right\}~.
\end{equation}

Since the function $F$ may depend on time through the function $K$, the Hamiltonian in the new coordinates is $H' = H + \frac{\partial F}{\partial t}$.
Imposing that the function $F$ generates a canonical transformation, is equivalent to demand that
\begin{align}
	\dot{q'}_i = & \frac{\partial H'}{\partial p_i} = \frac{\partial H}{\partial q_j} \frac{\partial q_j}{\partial p_i} + \frac{\partial H}{\partial p_{0,j}} \frac{\partial p_{0,j}}{\partial p_i} + \frac{\partial^2 K}{\partial p_{0,j} \partial t} \frac{\partial p_{0,j}}{\partial p_i}~, &
	\dot{p}_i = & - \frac{\partial H}{\partial q'_i} = - \frac{\partial H}{\partial q_j} \frac{\partial q_j}{\partial q'_i}~. \label{def:canonical}
\end{align}
We then find for the time derivatives
\begin{subequations}
\begin{align}
	\dot{q'}_i = & \left(\dot{q}_j + \frac{\partial^2 K}{\partial p_{0,j} \partial t}\right) \left\{ \delta_{ij} + \gamma \delta \sqrt{p_{0,k} p_{0,k}} \left( \delta_{ij} + \frac{p_{0,i} p_{0,j}}{p_{0,k} p_{0,k}} \right) - \gamma^2 p_{0,k} p_{0,k} \left[\epsilon \delta_{ij} + (2 \epsilon - \delta^2) \frac{p_{0,j} p_{0,i}}{ p_{0,k} p_{0,k}} \right] \right\} + \nonumber \\
	& + \left(q_j + \frac{\partial K}{\partial p_{0,j}}\right) \dot{p}_{0,m} \left\{ \frac{\gamma \delta}{\sqrt{p_{0,k} p_{0,k}}} \left( p_{0,m} \delta_{ij} + p_{0,i} \delta_{jm} + p_{0,j} \delta_{mi} - p_{0,m} \frac{p_{0,i} p_{0,j}}{p_{0,k} p_{0,k}} \right) + \right. \nonumber \\
	& \qquad \left. - \gamma^2 \left[ 2 \epsilon \left( p_{0,m} \delta_{ij} + p_{0,i} \delta_{jm} + p_{0,j} \delta_{mi} \right) - \delta^2 \left( p_{0,i} \delta_{jm} + p_{0,j} \delta_{mi} \right) \right] \right\}~,\\
	\dot{p}_i = & \dot{p}_{0,j} \left[\delta_{ij} - \gamma \delta \sqrt{p_{0,k} p_{0,k}} \left( \delta_{ij} + \frac{p_{0,i} p_{0,j}}{p_{0,k} p_{0,k}} \right) + \gamma^2 (\epsilon + \delta^2) p_{0,k} p_{0,k} \left(\delta_{ij} + 2 \frac{p_{0,i} p_{0,j}}{p_{0,k} p_{0,k}}\right) \right]~,
\end{align}
\end{subequations}
We will also need the following relations
\begin{subequations}
\begin{align}
	\frac{\partial p_{0,j}}{\partial p_i} 
	= & \delta_{ij} + \gamma\delta \sqrt{p_{0,k} p_{0,k}} \left(\delta_{ij} + \frac{p_{0,i} p_{0,j}}{p_{0,k} p_{0,k}}\right) - \gamma^2 p_{0,k} p_{0,k} \left[ \epsilon \delta_{ij} + (2 \epsilon - \delta^2) \frac{p_{0,i} p_{0,j}}{p_{0,k} p_{0,k}}\right]~,\\
	\frac{\partial q_j}{\partial p_i} 
%
	& - \frac{\partial^2 K}{\partial p_{0,j} \partial p_{0,k}} \frac{\partial p_{0,k}}{\partial p_i} = \\
	= & q'_l \left\{- \frac{\gamma \delta}{\sqrt{p_{0,k} p_{0,k}}} \left( p_{0,i} \delta_{jl} + p_{0,j} \delta_{li}  + p_{0,l} \delta_{ij} - \frac{p_{0,i} p_{0,j} p_{0,l}}{p_{0,k} p_{0,k}} \right) + \right. \nonumber \\
	& \left. \qquad + \gamma^2 \left[ 2 \epsilon ( p_{0,i} \delta_{jl} + p_{0,j} \delta_{li} + p_{0,l} \delta_{ij}) + \delta^2 ( p_{0,j} \delta_{li} + p_{0,l} \delta_{ij}) \right] \right\} + \nonumber \\
	& - \frac{\partial^2 K}{\partial p_{0,j} \partial p_{0,l}} \left\{ \delta_{il} + \gamma\delta \sqrt{p_{0,k} p_{0,k}} \left(\delta_{il} + \frac{p_{0,i} p_{0,l}}{p_{0,k} p_{0,k}}\right) - \gamma^2 p_{0,k} p_{0,k} \left[ \epsilon \delta_{il} + (2 \epsilon - \delta^2) \frac{p_{0,i} p_{0,l}}{p_{0,k} p_{0,k}}\right] \right\} =\\
	= & q_l \left\{ - \frac{\gamma \delta}{\sqrt{p_{0,k} p_{0,k}}} \left( p_{0,i} \delta_{jl} + p_{0,j} \delta_{li} + p_{0,l} \delta_{ij} - \frac{p_{0,i} p_{0,j} p_{0,l}}{p_{0,k} p_{0,k}} \right) + \right. \nonumber \\
	& \left. + \gamma^2 \left[ 2 \epsilon \left( p_{0,i} \delta_{jl} + p_{0,j} \delta_{li} + p_{0,l} \delta_{ij} \right) - \delta^2 \left( p_{0,i} \delta_{jl} + p_{0,l} \delta_{ji} \right) \right] \right\} + \nonumber \\
	& - \frac{\partial^2 K}{\partial p_{0,j} \partial p_{0,l}} \left\{ \delta_{il} + \gamma\delta \sqrt{p_{0,k} p_{0,k}} \left(\delta_{il} + \frac{p_{0,i} p_{0,l}}{p_{0,k} p_{0,k}}\right) - \gamma^2 p_{0,k} p_{0,k} \left[ \epsilon \delta_{il} + (2 \epsilon - \delta^2) \frac{p_{0,i} p_{0,l}}{p_{0,k} p_{0,k}}\right] \right\}~,\\
	\frac{\partial q_j}{\partial q'_i} = & \delta_{ij} - \gamma \delta \sqrt{p_{0,k} p_{0,k}} \left( \delta_{ij} + \frac{p_{0,i} p_{0,j}}{p_{0,k} p_{0,k}} \right) + \gamma^2 (\epsilon + \delta^2) p_{0,k} p_{0,k} \left(\delta_{ij} + 2 \frac{p_{0,i} p_{0,j}}{p_{0,k} p_{0,k}}\right)~.
\end{align}
\end{subequations}
Imposing the relations in \eqref{def:canonical}, we then find that $K$ depends only on time.
Notice that, since the function $K(t)$ depends only on time, it is completely uninfluencial for the equations of motion.
We can therefore safely pose $K=0$, obtaining the generating function in \eqref{eqn:generating_function}.

\end{appendix}


\begin{thebibliography}{10}

\bibitem{Gross1988_1}
D.~J. Gross and P.~F. Mende, ``{String theory beyond the Planck scale},'' {\em
  Nuclear Physics B}, vol.~303, pp.~407--454, jul 1988.

\bibitem{Amati1989_1}
D.~Amati, M.~Ciafaloni, and G.~Veneziano, ``{Can spacetime be probed below the
  string size?},'' {\em Phys. Lett. B}, vol.~216, no.~1, pp.~41--47, 1989.

\bibitem{Maggiore1993_1}
M.~Maggiore, ``{A generalized uncertainty principle in quantum gravity.},''
  {\em Phys. Lett. B}, vol.~304, no.~1, pp.~65--69, 1993.

\bibitem{Maggiore1993_2}
M.~Maggiore, ``{The algebraic structure of the generalized uncertainty
  principle.},'' {\em Phys. Lett. B}, vol.~319, no.~September, pp.~83--86,
  1993.

\bibitem{Garay1995_1}
L.~J. Garay, ``{Quantum Gravity and Minimum Length},'' {\em International
  Journal of Modern Physics A}, vol.~10, no.~02, pp.~145--165, 1995.

\bibitem{Scardigli1999_1}
F.~Scardigli, ``{Generalized uncertainty principle in quantum gravity from
  micro-black hole gedanken experiment},'' {\em Phys. Lett. B}, vol.~452,
  no.~1–2, pp.~39--44, 1999.

\bibitem{Capozziello2000}
S.~Capozziello, G.~Lambiase, and G.~Scarpetta, ``{Generalized Uncertainty
  Principle from Quantum Geometry},'' {\em International Journal of Theoretical
  Physics}, vol.~39, no.~1, pp.~15--22, 2000.

\bibitem{AmelinoCamelia2002_1}
G.~Amelino-Camelia, ``{Doubly-Special Relativity: First Results and Key Open
  Problems},'' {\em International Journal of Modern Physics D}, vol.~11,
  no.~10, pp.~1643--1669, 2002.
  
\bibitem{Kurkov}
M.~A.~Kurkov, F.~Lizzi, D.~Vassilevich, ``{High energy bosons do not propagate},'' {\em Phys. Lett. B}, vol.~731, pp.~311--315, 2014.

\bibitem{Kempf1995_1}
A.~Kempf, G.~Mangano, and R.~B. Mann, ``{Hilbert space representation of the
  minimal length uncertainty relation.},'' {\em Phys. Rev. D}, vol.~52,
  pp.~1108--1118, 1995.

\bibitem{Ali2011_1}
A.~F. Ali, S.~Das, and E.~C. Vagenas, ``{Proposal for testing quantum gravity
  in the lab},'' {\em Phys. Rev. D}, vol.~84, p.~44013, aug 2011.

\bibitem{Pramanik2013_1}
S.~Pramanik and S.~Ghosh, ``{GUP-Based and Snyder Noncommutative Algebras,
  Relativistic Particle Models, Deformed Symmetries and Interaction: a Unified
  Approach},'' {\em International Journal of Modern Physics A}, vol.~28,
  no.~27, p.~1350131, 2013.

\bibitem{Pramanik2014_1}
S.~Pramanik, S.~Ghosh, and P.~Pal, ``{Conformal invariance in noncommutative
  geometry and mutually interacting Snyder particles},'' {\em Phys. Rev. D},
  vol.~90, p.~105027, nov 2014.
  
\bibitem{Deriglazov}
A.~A.~Deriglazov, W.~G.~Ram\'irez, ``{Recent progress on the description of relativistic spin: vector model of spinning particle and rotating body with gravimagnetic moment in General Relativity},'' {\em Adv. Math. Phys.}, vol.~2017, p.~7397159, 2017.

\bibitem{Das2008_1}
S.~Das and E.~C. Vagenas, ``{Universality of Quantum Gravity Corrections},''
  {\em Phys. Rev. Lett.}, vol.~101, p.~221301, nov 2008.

\bibitem{Bosso2017a}
P.~Bosso, S.~Das, and R.~B. Mann, ``{Planck scale corrections to the harmonic
  oscillator, coherent, and squeezed states},'' {\em Phys. Rev. D},
  vol.~96, p.~066008, sep 2017.

\bibitem{Gusson2018}
M.~F. Gusson, A.~O.~O. Gon{\c{c}}alves, R.~O. Francisco, R.~G. Furtado, J.~C.
  Fabris, and J.~A. Nogueira, ``{Dirac $\delta$-function potential in
  quasiposition representation of a minimal-length scenario},'' {\em The
  European Physical Journal C}, vol.~78, p.~179, mar 2018.

\bibitem{Kober2010}
M.~Kober, ``{Gauge theories under incorporation of a generalized uncertainty
  principle},'' {\em Phys. Rev. D}, vol.~82, no.~8, pp.~1--12, 2010.

\bibitem{Das2016}
P.~Das, S.~Pramanik, and S.~Ghosh, ``{Particle on a torus knot: Constrained
  dynamics and semi-classical quantization in a magnetic field},'' {\em Annals
  of Physics}, vol.~374, pp.~67--83, nov 2016.

\bibitem{Das2016a}
P.~Das and S.~Ghosh, ``{Noncommutative geometry and fluid dynamics},'' {\em The
  European Physical Journal C}, vol.~76, p.~627, nov 2016.

\bibitem{Goldstein_H}
H.~Goldstein, {\em {Classical Mechanics}}.
\newblock A-W series in advanced physics, Addison-Wesley Press, 1950.

\bibitem{Das2009_1}
S.~Das and E.~C. Vagenas, ``{Phenomenological implications of the generalized
  uncertainty principle},'' {\em Canadian Journal of Physics}, vol.~87, no.~3,
  pp.~233--240, 2009.
  
\bibitem{Galan2007}
P.~Gal\'an, G.~A.~Mena~Marug\'an, ``{Canonical Realizations of Doubly Special Relativity},'' {\em International Journal of Modern Physics D}, vol.~16, no.~7, pp.~1133-1147, 2007.
  
\bibitem{AmelinoCamelia2013_1}
G.~Amelino-Camelia, ``{Challange to Macroscopic Probes of Quantum Spacetime Based on Noncommutative Geometry},'' {\em Phys. Rev. Lett.}, vol.~111, no.~10 pp.~101301, 2013.

\end{thebibliography}
\end{document}